%% file: km3net_grb221009A.tex
\documentclass[a4paper,11pt]{article}
\usepackage{jcappub} 

\usepackage{multirow}

\arxivnumber{2404.05354} 
\title{\boldmath Search for Neutrino Emission from GRB 221009A using the KM3NeT ARCA and ORCA detectors}

\collaboration{\includegraphics[height=17mm]{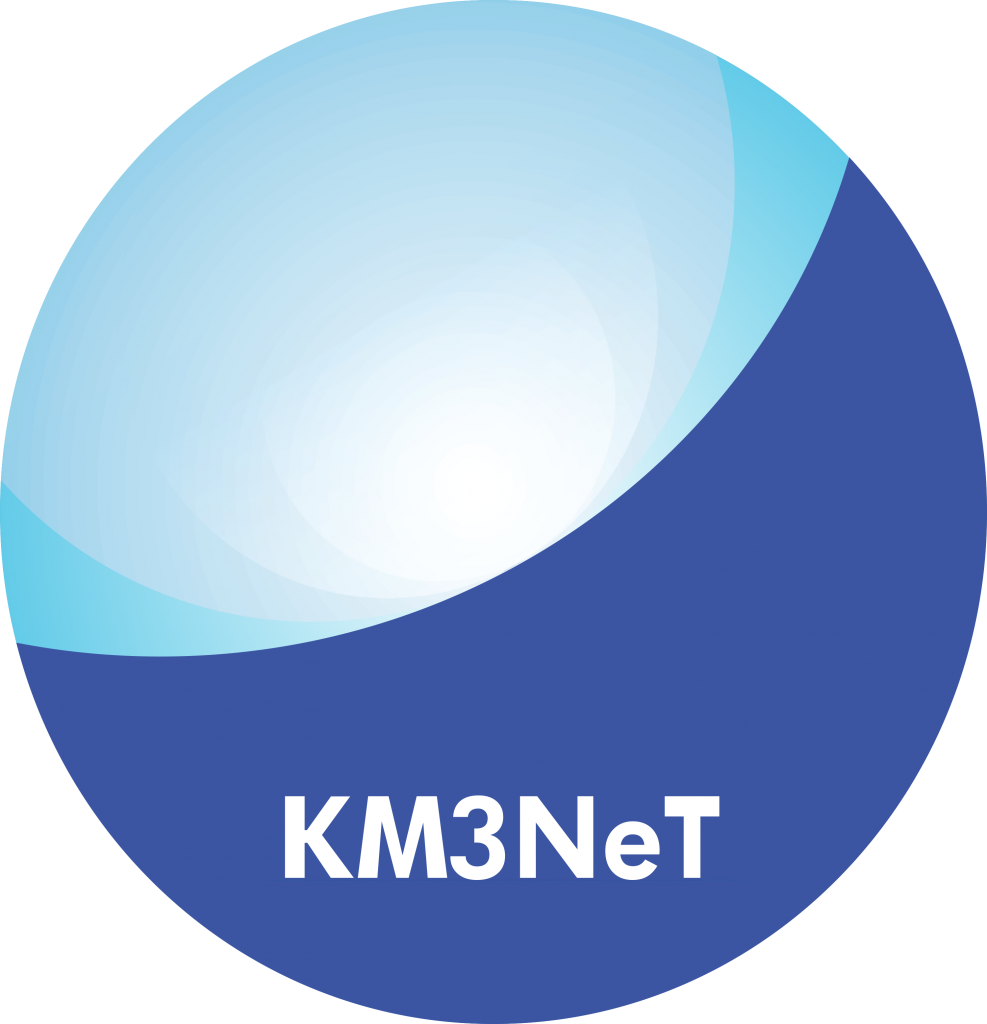}\\[6pt]
  The KM3NeT collaboration}

\input{KM3NeT_authorlist_GRB221009A_JCAP.tex}

\emailAdd{Juan.Palacios@ific.uv.es}
\emailAdd{lestum@cppm.in2p3.fr}
\emailAdd{vannoye@cppm.in2p3.fr}
\emailAdd{km3net-pc@km3net.de}

\abstract{Gamma-ray bursts are promising candidate sources of high-energy astrophysical neutrinos. The recent GRB 221009A event, identified as the brightest gamma-ray burst ever detected, provides a unique opportunity to investigate hadronic emissions involving neutrinos. The KM3NeT undersea neutrino detectors participated in the worldwide follow-up effort triggered by the event, searching for neutrino events. In this paper, we summarize subsequent searches, in a wide energy range from MeV up to a few PeVs. No neutrino events are found in any of the searches performed. Upper limits on the neutrino emission associated with GRB 221009A are computed.}

\begin{document}
\maketitle
\flushbottom

\section{Introduction: GRB 221009A} \label{section_intro}
On October 9, 2022, at 13:16:59.0 UT ($T_0$ hereafter) the Fermi satellite detected an extraordinarily bright transient phenomenon by the Gamma-Ray Burst Monitor (GBM)~\cite{2022GCN.32636....1V}. Shortly after that, at 14:10:17 UT, the Swift Burst Alert Telescope (BAT) detected with better angular accuracy a transient event consistent with the location of Fermi-GBM, with additional detection of candidate counterparts by Swift-XRT and Swift-UVOT~\cite{2022GCN.32632....1D}. This transient phenomenon was identified as a long Gamma-Ray Burst (GRB), whose most likely origin is a supernova from the core collapse of a supermassive star~\cite{2022GCN.32686....1C}.

The light curve measured by Fermi-GBM is composed of two consecutive emission periods: a first single isolated peak followed by a longer multi-pulsed episode in the \mbox{10--1000\,keV} energy range. The duration corresponding to 90\% of the central emission from this GRB ($T_{90}$  hereafter) is around $327$\,s~\cite{2022GCN.32642....1L}. As the exceptional brightness of this event saturated the Fermi GBM and LAT instruments during the main emission period, this value may be an overestimation of the real $T_{90}$~\cite{2023ApJ...952L..42L}. 

The most accurate location for GRB 221009A has been provided by the Swift spacecraft at R.A.(J2000) $=$ 288.26452$^\circ$ and DEC.(J2000) $=$ $+$ 19.77350$^\circ$, with a $90\%$ confidence error radius of $0.61$ arcseconds~\cite{2022GCN.32632....1D}. This location has been supported by the observations of multiple observatories~[\cite{2022GCN.32660....1G,2022GCN.32668....1F,2022GCN.32641....1S}]. The afterglow phase was detected and characterized in a wide frequency range, with X-ray~\cite{2022GCN.32680....1T}, optical~\cite{2022GCN.32638....1P,2023ApJ...946L..22F}, and radio \cite{2022GCN.32653....1B,2022GCN.32757....1L} observations, in one of the largest follow-up campaigns ever.

The LHAASO Water Cherenkov Detector Array detected photons from $200$\,GeV up to energies above $\,10$\,TeV, the highest energy observed so far from a GRB~\cite{2023Sci...380.1390L,2023SciA....9J2778C}. Additionally, the Fermi-LAT satellite detected photons with energies close to 99\,GeV, which is to date the highest-energy detection by this instrument for a GRB~\cite{2022GCN.32658....1P}. A GRB as extraordinary as GRB 221009A is expected to occur only once every ten thousand years~\cite{2023ApJ...946L..31B}.

The exceptional intensity of this GRB may be explained by its proximity, together with the fact that the jet emission is considered to be very collimated~\cite{2023ApJ...946L..23L}. A dedicated study suggests that the broadband afterglow emission can indeed be explained by a broad structured jet~\cite{2023SciA....9I1405O}. The redshift has been estimated to be $z=0.151$, following the optical observations of the X-shooter VLT and the 10.4\,m GTC telescopes~\cite{2022GCN.32648....1D,2022GCN.32686....1C}. This corresponds to a luminosity distance $d_L=745$\,Mpc~\cite{2023arXiv230207891M}. The isotropic-equivalent energy and luminosity are estimated to be $10^{55}$\,erg and $9.9\times10^{53}$\,erg\,s$^{-1}$ respectively, which are among the highest values known for a GRB to date~\cite{2023ApJ...952L..42L}.

\subsection{Neutrino emission from GRB 221009A} 

High-energy neutrinos (in the TeV--PeV energy range) are expected to be generated through a wide variety of astrophysical processes in GRB events. This is of particular interest in the case of GRB 221009A, as a recent study~\cite{2023arXiv231011821W} hinted towards a significant contribution of hadronic processes in its multi-wavelength emission.

Current models of the prompt emission from a GRB describe it as a high-temperature compacted plasma whose kinetic energy can be released through internal shocks. These shocks can accelerate protons and other atomic nuclei from the relativistic outflow, which may produce TeV--PeV neutrinos through photo-meson production from the interaction with surrounding photon fields~\cite{PhysRevLett.78.2292}. Currently, this is expected to be the main mechanism for neutrino production~\cite{2022arXiv220206480K}. Nevertheless, there are still many other models proposed that could explain neutrino emissions from GRBs such as subphotospheric neutrino production~\cite{2013PhRvL.111m1102M}, neutrino eruptions in jet propagation processes~\cite{2014ApJ...790...59X}, regular core-collapse processes~\cite{2013ApJ...778...81K} or even detectable neutrinos from neutron collisions~\cite{2022ApJ...941L..10M}.

The KM3NeT Collaboration performed a real-time search for neutrino emission, reporting a non-detection of candidate events during the [$T_0-50$\,s, $T_0+5000$\,s] time window~\cite{2022GCN.32741....1K}.

The IceCube Collaboration also reported a non-detection of neutrinos coming from GRB 221009A ~\cite{2022GCN.32665....1I}. The real-time analyses, conducted using the time windows [$T_0$ $-$ $1$\,hr, $T_0$ $+$ $2$\,hr] and $T_0$ $\pm$ $1$\,day, were based on a Fast Response Analysis (FRA) framework optimized for TeV--PeV neutrino detection. Refined upper limits in the time-integrated energy-scaled neutrino emission from GRB 221009A were further reported~\cite{2023ApJ...946L..26A}. The Baikal-GVD Collaboration also performed an independent neutrino search with no significant detection~\cite{2023arXiv230813829A}.

In this paper, the results of refined searches in the KM3NeT detectors for a neutrino signal compatible with the location of GRB 221009A are shown, covering various time windows. The data were analyzed with improved calibrations with respect to the real-time search and considering systematic effects such as the angular uncertainty in the reconstruction of the events.

\section{The KM3NeT neutrino detectors} \label{section_km3net}

KM3NeT~\cite{2016JPhG...43h4001A} is an international research collaboration currently deploying two deep-sea neutrino telescopes in the Mediterranean Sea: ARCA (Astroparticle Research with Cosmics in the Abyss) and ORCA (Oscillation Research with Cosmics in the Abyss). These detectors consist of three-dimensional arrays capable of detecting the Cherenkov light emission induced by the motion of relativistic charged particles resulting from neutrino interactions.

The active component of the KM3NeT detectors is the Digital Optical Module (DOM), a pressure-resistant glass sphere housing 31 Photomultiplier Tubes (PMTs)~\cite{Aiello_2022}. The DOMs are embedded in vertical strings called Detection Units (DUs), with 18 DOMs incorporated in each DU.

ORCA is located 40\,km from Toulon (France) at a depth of 2.5\,km. The distance between DOMs is optimized for detecting neutrino events in the GeV energy range. Although the main physics goal of ORCA is the study of neutrino oscillation properties by the detection of atmospheric neutrinos, the energy range covered makes it a well-suited instrument to perform GeV neutrino astronomy as well.

ARCA, a larger array that will instrument about 1\,km$^3$ of deep-sea water, is located 100\,km offshore from Capo Passero (Sicily, Italy) at a depth of 3.5\,km. The separation between DOMs makes this detector sensitive to neutrino events from sub-TeV energies up to a few PeV. The main physics goal of ARCA is the study of high-energy neutrinos originating from astrophysical sources. MeV neutrinos can also be detected in ARCA and ORCA by looking for an increase in the rate of PMT coincidences in each DOM~\cite{KM3NeT:2021moe,KM3NeT:2021oaa}.

The high duty cycle of the detectors above 95\%, the energy coverage from MeV to PeV, the full-sky coverage and the angular resolution below one degree for energies above 10\,TeV for the searches performed in this work, turn the KM3NeT detectors into suitable instruments to perform multi-messenger studies.

The final detector configurations will comprise 115~DUs for ORCA and 230~DUs for ARCA. When GRB 221009A took place, ARCA was operated with 21~DUs, while ORCA had 10~DUs in operation.

\section{Search methods and results} \label{section_results}

The methods and results of the different refined searches performed using the KM3NeT detectors are presented in this section. A description of the analysis method employed in the GeV--PeV energy range is provided in Section~\ref{section_search_GeV_PeV}. The results for the TeV--PeV analyses, using ARCA data, are presented in Section~\ref{section_TeV_PeV_neutrinos}. The GeV--TeV neutrino searches, using ORCA data, are detailed in Section~\ref{section_GeV_TeV_neutrinos}. Finally, the method and results in the MeV range are presented in Section~\ref{subsection_MeV_neutrinos}.

\subsection{Search method: GeV--PeV neutrinos} \label{section_search_GeV_PeV}

\begin{figure}[t]
    \centering
    \includegraphics[width=0.6\textwidth]{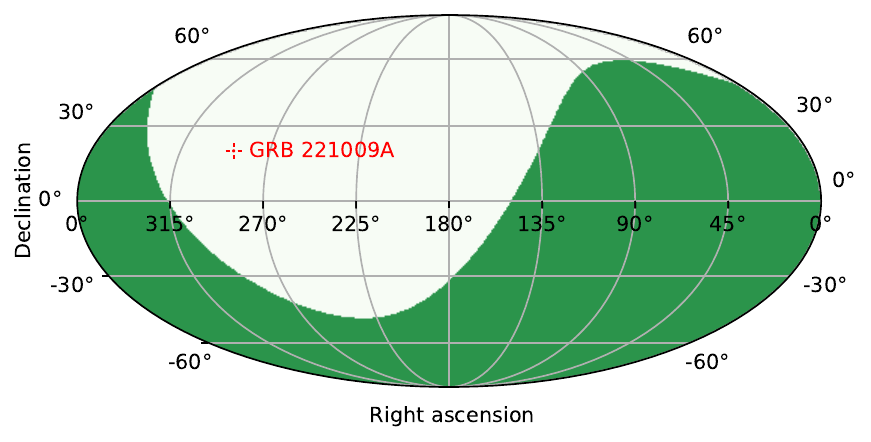}
     \caption{Skymap with the position of GRB 221009A in equatorial coordinates (J2000). The green shadowed region denotes the instantaneous visibility for upgoing events of ARCA at $T_0$. In the case of ORCA, the visibility region is similar.} 
     \label{fig_ARCA21_visibility}
 \end{figure}

At $T_0$, the location of GRB 221009A was above the horizon in the local sky of the KM3NeT detectors, as shown in Figure \ref{fig_ARCA21_visibility}. Therefore, neutrinos coming from the direction of this source are reconstructed as downgoing events. The downgoing data sample used in this analysis is composed of events reconstructed with a cosine of the zenith angle larger than $0.1$. This part of the sky is dominated by atmospheric muon events, and strict event selections must be applied to reduce this background.

Two time windows, shorter than 24 hours, are considered: one for [$T_0$, $T_{90}$], and another longer one for [$T_0$ $-$ $50$\,s, $T_0$ $+$ $5000$\,s], the latter covering the high-energy emission seen by Fermi-LAT~\cite{2022GCN.32658....1P} and LHAASO~\cite{2023Sci...380.1390L}. In these two time windows, the GRB remains above the horizon. Furthermore, two searches are conducted in the $T_0\pm1$ day range: one using only events reconstructed as upgoing, and another one employing events reconstructed as downgoing. The search using upgoing events is motivated by the fact that the location of GRB 221009A in the local detector sky is below the horizon for about 45\% of one day.

For events with reconstructed energy above a few GeV in ORCA and a few hundred GeV in ARCA, two main topologies can be identified: track-like events, where the Cherenkov PMT pulses (\textit{hits}) are compatible with a straight line, and cascade-like events, where the spatial distribution of the hits is compatible with a quasi-spherical light emission. These topologies are associated with the flavor of the interacting neutrino and with the kind of interaction. Track-like signatures primarily arise from the charged-current interactions of muon neutrinos and some charged-current interactions of tau neutrinos. Cascade-like topologies arise from the interactions of the neutrinos of the remaining flavors and neutral current muon neutrino interactions~\cite{2024arXiv240208363V}. In the analyses described in this paper, only track-like events are considered, because of their superior angular resolution.

The search method is based on a binned ON/OFF technique~\cite{1983ApJ...272..317L}. The ON region is defined as the region of the sky where a signal is expected. The OFF region corresponds to a fraction of the sky where the background level is comparable to the ON region but only background events are expected. As ON region, a circular area centered on the GRB position, i.e. the Region of Interest (RoI) is used. For the OFF region, declination bands are considered for searches over time windows longer than one day, where the atmospheric background depends only on the declination due to the Earth's rotation. For shorter time windows, elevation bands are used as the OFF region to account for the dependence of the background on the local sky. In both cases, the number of events in the OFF region is re-scaled to the solid angle and to the time covered by the ON region.

The event selections are determined to obtain a background level such that the detection of one event in the ON region is sufficient to obtain a $3\sigma$ excess above the background-only hypothesis. Using a $2$-sided convention, this corresponds to an expected background lower or equal to $2.7 \times 10^{-3}$ events. This optimization takes into account the performance of the detector through two instrument response functions: the Point Spread Function (PSF) and the detector effective area.

The PSF describes the dispersion of a signal coming from the direction of the GRB due to the angular resolution of the detector. Systematic uncertainties are included in the PSF, mainly coming from the determination of the absolute orientation of the detectors and the geometrical shape of the detection units as reconstructed through acoustic triangulation methodologies. The resulting median of the PSF for the data sample considered in this work is estimated at $0.8^\circ$ for ARCA and at 1.2$^\circ$ for ORCA, for an $E^{-2}$ neutrino spectrum.
 
The detector effective area $A_{\rm eff}^{\delta}(E)$ at a declination $\delta$ is defined as the quantity that provides the number of expected signal events $N_{\rm s}$ when convoluted with the differential neutrino flux $\Phi (E)$,
\begin{equation}
    N_{\rm s} = \int dt \int dE \cdot \Phi (E) \cdot A_{\rm eff}^{\delta}(E).
\end{equation}
The instrument response functions are computed using dedicated Monte Carlo simulations of neutrino interactions generated with the gSeaGen software~\cite{2020CoPhC.25607477A} in the energy range between 1\,GeV and 10\,TeV for ORCA and from 100\,GeV to 100\,PeV for ARCA. The systematic uncertainty on the detector effective area is conservatively estimated to be 30\%, mainly due to the uncertainties on the properties of the seawater medium and of the PMT, such as the light absorption length or the PMT quantum efficiency~\cite{2024arXiv240311946T}.

A differential neutrino flux $\Phi (E) = \Phi_0 (E/E_0)^{-\gamma}$, with spectral index $\gamma=2$ and $E_0=1$\,GeV, is considered in all searches. The normalization factor $\Phi_0$ is left as a free parameter in order to determine its value or an upper limit in the non-detection case. The shape assumed for the neutrino spectrum corresponds to the one expected from hadrons accelerated by a second-order Fermi mechanism in strong shocks~\cite{1978ApJ...221L..29B,1978MNRAS.182..147B}.

\subsection{Results of the TeV--PeV neutrino searches with the ARCA detector} \label{section_TeV_PeV_neutrinos}

A dataset with approximately 70 days of livetime is used to estimate the expected background in the ARCA searches. The selection criteria are based on the quality of the event reconstruction, the estimated angular uncertainty of the event, the number of hits used in the reconstruction, and the estimated length of the reconstructed track. In the analyses of downgoing tracks the estimated energy of the events is also considered. The results for all searches are summarized in Table \ref{table_main_results}.

\begin{table}[t]
\centering
\fontsize{6.5}{11}\selectfont
\begin{tabular}{cccccccc}
\hline \hline \textsc{SEARCH}    & \multicolumn{7}{c}{\textsc{KM3NeT $90\%$~CL upper limits on neutrino emission from GRB 221009A} }     \\ \hline \hline
\multicolumn{1}{c|}{\multirow{2}{*}{\begin{tabular}[c]{@{}c@{}} \\ \textbf{ARCA}\\ \textbf{(TeV--PeV)}\end{tabular}}} & \multicolumn{7}{c}{Results for neutrino flux $\Phi (E) = \Phi_0 (E/E_0)^{-2}$ at $E_0=1$\,GeV}      \\
\multicolumn{1}{c|}{}                                                                        & \begin{tabular}[c]{@{}c@{}} RoI \\ radius\end{tabular}        & \begin{tabular}[c]{@{}c@{}} Expected \\ background \\ $(\times\, 10^{-3})$ \end{tabular} & \begin{tabular}[c]{@{}c@{}} $\Phi_0$ UL \\ {[}GeV$^{-1}$\,cm$^{-2}$\,s$^{-1}${]} \end{tabular} & \begin{tabular}[c]{@{}c@{}} $E_{\rm min}$ \\ {[}TeV{]} \end{tabular} & \begin{tabular}[c]{@{}c@{}} $E_{\rm max}$ \\ {[}PeV{]} \end{tabular} & \begin{tabular}[c]{@{}c@{}} Fluence $\mathcal{F}$ UL \\ {[}GeV\,cm$^{-2}${]}\end{tabular} & \begin{tabular}[c]{@{}c@{}} $E^2F(E)$ UL \\ {[}GeV\,cm$^{-2}${]}\end{tabular}   \\ \hline
\multicolumn{1}{c|}{$T_{90}$}                                                                     & $2.1^\circ$ & $2.64\pm0.02$ & $2.5\times 10^{-3} $& $34$ &   $13$ &   4.9 &   $0.83$  \\
\multicolumn{1}{c|}{$T_0$[$-50$\,s, $+5000$\,s]}    & $1.1^\circ$ &  $2.53\pm0.04$ & $2.8\times  10^{-4} $ & $110$ &   $27$ & 7.9 &  1.4   \\
\multicolumn{1}{c|}{$T_0\pm$1d downgoing}   & $1.0^\circ$ &    $2.6\pm0.1$& $2.5\times 10^{-5}$ & $220$ &    $36$    &  22 & 4.4  \\
\multicolumn{1}{c|}{$T_0\pm$1d upgoing}  & $1.7^\circ$ & $2.7\pm0.2$ & $6.2\times 10^{-6} $& $8.1$ &  $7.7$    &   7.4 & 1.1  \\ \hline \hline
\multicolumn{1}{c|}{\multirow{2}{*}{\begin{tabular}[c]{@{}c@{}} \\ \textbf{ORCA}\\ \textbf{(GeV--TeV)}\end{tabular}}}                      & \multicolumn{7}{c}{Results for neutrino flux $\Phi (E) = \Phi_0 (E/E_0)^{-2}$ at $E_0=1$\,GeV}           \\ \multicolumn{1}{c|}{}                                                                        & \begin{tabular}[c]{@{}c@{}} RoI \\ radius\end{tabular}        & \begin{tabular}[c]{@{}c@{}} Expected \\ background \\ $(\times\, 10^{-3})$ \end{tabular} & \begin{tabular}[c]{@{}c@{}} $\Phi_0$ UL \\ {[}GeV$^{-1}$\,cm$^{-2}$\,s$^{-1}${]} \end{tabular} & \begin{tabular}[c]{@{}c@{}} $E_{\rm min}$ \\ {[}GeV{]} \end{tabular} & \begin{tabular}[c]{@{}c@{}} $E_{\rm max}$ \\ {[}TeV{]} \end{tabular} & \begin{tabular}[c]{@{}c@{}} Fluence $\mathcal{F}$ UL \\ {[}GeV\,cm$^{-2}${]}\end{tabular} & \begin{tabular}[c]{@{}c@{}} $E^2F(E)$ UL \\ {[}GeV\,cm$^{-2}${]}\end{tabular}   \\ \hline
\multicolumn{1}{c|}{$T_{90}$}  & $2.0^\circ$ & $2.61\pm0.04$ & $14$ & $150$ &  $9.1$ &  $1.9\times 10^{4} $&  $4.5\times 10^{3}$ \\
\multicolumn{1}{c|}{$T_0$[$-50$\,s, $+5000$\,s]}  & $5.4^\circ$ &  $2.6\pm0.2$ & 1.9 & $54$ &  $8.7$ &   $4.9\times 10^{4}$   &  $9.6\times 10^{3}$ \\
\multicolumn{1}{c|}{$T_0\pm$1d downgoing} & $1.0^\circ$ &    $2.7\pm0.3$ & $1.0 \times 10^{-2} $& $68$ &  $8.8$  &  $8.5\times 10^{3}$  &  $1.7\times 10^{3}$ \\
\multicolumn{1}{c|}{$T_0\pm$1d upgoing}  & $1.2^\circ$ &  $2.7\pm0.3$ & $4.7\times 10^{-4}$& $130$ & $9.8$ &  $3.5\times 10^{2} $& $81$  \\ \hline \hline
\multicolumn{1}{c|}{\multirow{2}{*}{ \begin{tabular}[c]{@{}c@{}} \\  \textbf{MeV} \\ \textbf{search} \end{tabular}}}                                                    & \multicolumn{7}{c}{Results for quasi-thermal neutrino flux $F_{\bar{\nu}_e}(E) \propto E^2 \text{exp} (-3E/\langle E \rangle )$ at $\langle E \rangle=15$\,MeV}      \\
\multicolumn{1}{c|}{}                                                                        & \begin{tabular}[c]{@{}c@{}}Maximum\\ coincidence\\ level\end{tabular} & \begin{tabular}[c]{@{}c@{}}Expected\\ background\end{tabular}     & $p$-value                                               & \begin{tabular}[c]{@{}c@{}} $E_{\rm min}$ \\ {[}MeV{]} \end{tabular}   & \begin{tabular}[c]{@{}c@{}} $E_{\rm max}$ \\ {[}MeV{]} \end{tabular}    & \begin{tabular}[c]{@{}c@{}} Total \(\bar{\nu}_e\) flux \\ {[}cm$^{-2}${]} \end{tabular}   & \begin{tabular}[c]{@{}c@{}} \(E_{\rm tot,\nu}^{{\rm iso,} 90\%}\) \\ {[}erg{]} \end{tabular} \\ \hline
\multicolumn{1}{c|}{$T_{90}$}    & 27    & 29    & 0.99    & \multirow{2}{*}{5}                                   & \multirow{2}{*}{30}                                   & $2.5 \times 10^{9}$  & $5.1 \times 10^{62}$        \\ \multicolumn{1}{c|}{$T_0$[$-50$\,s, $+5000$\,s]}       & 32   & 33   & 0.79   &     &       & $4.8 \times 10^{9}$  &  $9.7 \times 10^{62}$     \\ \hline \hline
\end{tabular}
\caption{$90\%$~CL upper limits on the neutrino emission from GRB 221009A for the different time windows studied in each data sample. $E_{\rm min}$ and $E_{\rm max}$ correspond respectively to the 5\% and 95\% quantiles in the energy range of the expected neutrino flux for each search. For the GeV to PeV analyses, the searches for the $T_{90}$ and $T_0$[$-50$\,s, $+5000$\,s] time windows are only downgoing. The expected background level for each search is also presented.}
\label{table_main_results}
\end{table}

For the $T_0$ $\pm$ $1$ day searches, the RoI radius obtained for the upgoing search is larger than in the case of the downgoing study, as expected due to the lower atmospheric background. In the searches for downgoing neutrino candidates, the RoI radius increases as the time window duration decreases, which is expected since the use of shorter windows reduces the background. 

No candidate neutrino event coming from the ON region around GRB 221009A direction is found in any of the time windows. Upper limits (UL) on the neutrino emission from GRB 221009A are evaluated, taking into account the instrument response functions of the current detector configurations. 

The upper limit on the flux normalization factor $\Phi_0$ is determined as
\begin{equation}
    \Phi_0^{\rm UL} = \frac{\mu_{90}^{\rm FC}(n_b)}{\int dt \int dE \cdot \left( E/E_0 \right)^{-\gamma}\cdot A_{\rm eff}^{\delta} (E)},
\end{equation}
where the quantity $\mu_{90}^{\rm FC}(n_b)$ denotes the 90\% confidence level upper limit on the number of events according to the Feldman-Cousins method~\cite{1998PhRvD..57.3873F}, and $A_{\rm eff}^{\delta} (E)$ is the detector effective area at the declination of GRB 221009A.

The upper limit on the radiant fluence, defined as the per-flavor neutrino flux integrated in energy and time, over a given search time window is 
\begin{equation}
    \mathcal{F}^{\rm UL} = \Delta T \int_{E_{\rm min}}^{E_{\rm max}} dE \cdot E \cdot \Phi_0 ^{\rm UL} \cdot \left( \frac{E}{E_0} \right)^{-\gamma}, 
\end{equation}
where $\Delta T$ is the emission period considered, and $E_{\rm min}$ and $E_{\rm max}$ correspond to the 5\% and 95\% quantiles of the energy range for the neutrino flux~\cite{2023JCAP...08..072A}. 

The upper limit on the energy-scaled time-integrated neutrino flux, defined as $E^2F(E)=\Delta T \cdot \Phi_0^{UL} \cdot E_0^2$, was computed in order to perform a comparison with the upper limits reported by the \mbox{IceCube} Collaboration~\cite{2023ApJ...946L..26A}.

\subsection{Results of the GeV--TeV neutrino searches with the ORCA detector} \label{section_GeV_TeV_neutrinos}

A dataset with $\sim\,$41 days of livetime is used in the ORCA searches to estimate the expected number of background events. The same four time windows as for the ARCA analyses are inspected. The main difference comes from the use of a machine learning classification algorithm~\cite{2016arXiv160302754C} in the event selection, which aims to reject atmospheric muons. The classifier is introduced considering the larger impact of the atmospheric muon background in the ORCA analyses, due to the lower depth of the site. The event selection optimization procedure was applied to determine the best classification score and RoI radius for each considered time window.

No candidate neutrino event is found in coincidence with the position of GRB 221009A in any of the searches. As in the case of ARCA, upper limits on the neutrino flux normalization factor, time-integrated energy-scaled flux, and time-integrated energy-integrated flux are determined. The results are provided in Table \ref{table_main_results}. 

\subsection{Method and results for MeV neutrinos} \label{subsection_MeV_neutrinos}

The KM3NeT detectors can identify a burst of electron anti-neutrinos in the 5--30\,MeV energy range through the observation of a global increase in the rate of coincidences between PMTs in single DOMs. This method was developed to look for neutrinos from core-collapse supernovae~\cite{KM3NeT:2021moe,KM3NeT:2021oaa}. A coincidence is defined as at least four hits (PMT voltage above a given threshold)  within 10\,ns inside single DOMs for PMTs within 90~degrees of each other. Coincidences are aggregated in sliding windows of 500\,ms computed every 100\,ms independently for each detector before being summed together. This number of coincidences is referred to as the coincidence level.

The search for MeV neutrinos is made considering two time windows: [$T_0$, $T_{90}$] and [$T_0-50$\,s, $T_0+5000$\,s]. To compute the significance and associated p-value, the maximum coincidence level is evaluated for each time window and compared to the expected background, obtained using 30 days of data around $T_0$.

The 90\% confidence level upper limits on the number of coincidences due to a neutrino signal is computed following the Feldman-Cousins method~\cite{Feldman:1997qc}. This quantity is used to compute the upper limit on the total time-integrated neutrino flux and on the total energy emitted in isotropically distributed MeV neutrinos by the source $E_{\rm tot,\nu}^{{\rm iso,} 90\%}$ taking into account the inferred GRB distance. These upper limits are computed assuming a quasi-thermal emission of electron anti-neutrinos. The results are presented in Table~\ref{table_main_results}.

\section{Discussion} \label{section_discussion}

\begin{figure}[t]
    \centering
    \includegraphics[width=1.0\textwidth]{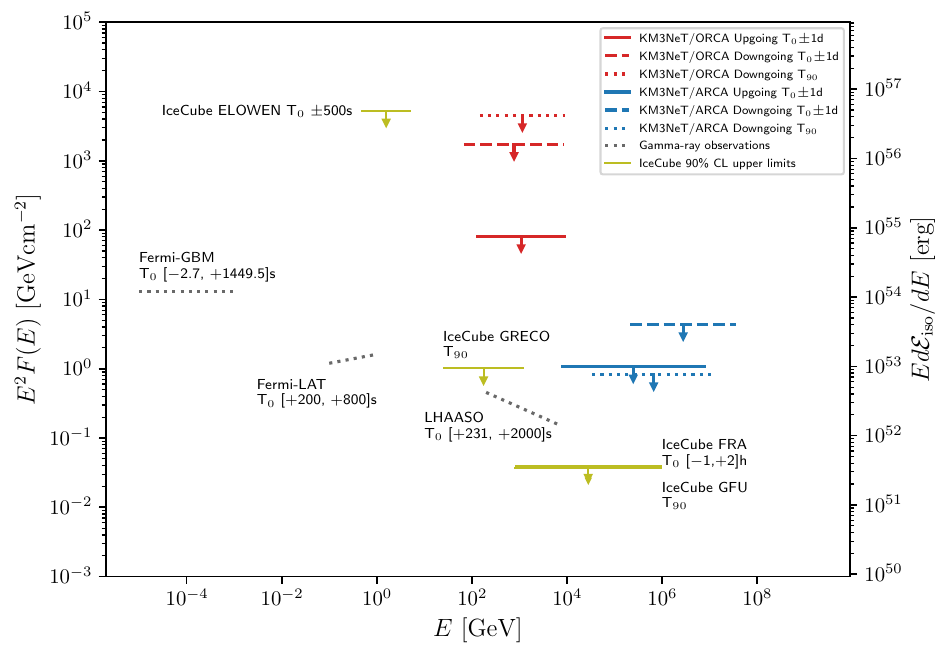}
     \caption{ $90\%$~CL upper limits on $E^2F(E)$, the energy-scaled time-integrated per flavor neutrino flux from GRB 221009A, for KM3NeT/ORCA (red lines) and KM3NeT/ARCA (blue lines). The results from IceCube, taken from~\cite{2023ApJ...946L..26A}, are also shown (green lines). Only the results for the $T_0\pm 1$ day and $T_{90}$ searches are included, using a neutrino spectral index $\gamma=2$, as they are the most relevant ones. For visualization purposes, the gamma-ray observations are also included (gray dashed lines), from Fermi-GBM~\cite{2023ApJ...952L..42L} (section 8), Fermi-LAT~\cite{2022GCN.32658....1P}, and LHAASO~\cite{2023Sci...380.1390L} (Table S2, assuming an intrinsic spectrum and standard EBL). The right axis indicates the differential isotropic equivalent energy.}   
     \label{fig_sensitivity_plot}
 \end{figure}

In Figure~\ref{fig_sensitivity_plot} the main results of this work are presented and compared with the upper limits derived by the IceCube Collaboration~\cite{2023ApJ...946L..26A}.  As a reference the observations of the photon flux by Fermi-GBM~\cite{2023ApJ...952L..42L}, Fermi-LAT~\cite{2022GCN.32658....1P}, and LHAASO~\cite{2023Sci...380.1390L} are included.

The right $y$-axis of Figure~\ref{fig_sensitivity_plot} denotes the differential isotropic equivalent energy $Ed\mathcal{E}_{\rm iso}/dE$. The upper limit on $E^2F(E)$ is translated into an upper limit on $Ed\mathcal{E}_{\rm iso}/dE$ using the relation $E^2F(E)=Ed\mathcal{E}_{\rm iso}/dE \times (1+z)/(4\pi d_L^2)$, accounting for the redshift $z$ and the luminosity distance $d_L$~\cite{2023ApJ...946L..26A}. 

In the case of the ARCA detector, the most restrictive upper limit on the neutrino flux normalization $\Phi_0$ is provided by the searches for upgoing neutrino candidates, see Table~\ref{table_main_results}. This can be explained by considering the lower atmospheric muon contamination in the upgoing sky. Instead, for the case of $E^2F(E)$ the most restrictive value is provided by the $T_{90}$ time window, as shown in Figure~\ref{fig_sensitivity_plot}. This is expected since $E^2F(E)$ is a time-integrated quantity, which favors shorter time window searches. 

Although the same argument applies in ORCA, due to the larger atmospheric muon contamination in the ORCA energy range, the performance of the downgoing searches for this detector is a factor 20 worse than for the upgoing sky, as can be seen in Table~\ref{table_main_results}.

The limits presented in this paper are not as restrictive as the ones provided by the IceCube Collaboration. This can be explained by considering the position of the GRB in the local sky of the KM3NeT detectors, which happened to be above the horizon at the time of the event, in contrast to the case of IceCube. Furthermore, the current partial configuration ($\sim$\,10\% of the full configuration) of the ARCA and ORCA telescopes results in a limited effective area.

\section{Conclusions} \label{section_conclusions}

Searches for a neutrino signal coming from GRB 221009A have been performed with the KM3NeT ARCA and ORCA detectors. No candidate neutrino events were found in coincidence with the GRB location. Taking into account the effective area of the current detector configurations, upper limits on the neutrino emission were obtained. Despite the reduced size of the detectors at the time of GRB 221009A, about 10\% of the final KM3NeT configuration, the results of these analyses can be used to provide constraints on different neutrino emission models for the case of GRB 221009A.

The ongoing construction of KM3NeT will increase the sensitivity for cosmic neutrino detection in the coming years by one order of magnitude at least, enhancing the relevant impact on forthcoming searches for neutrinos originating from GRBs. For a significant detection it is crucial to continuously monitor the sky with full coverage, using events coming from both the upgoing and downgoing sky regions. This can be achieved by exploiting the complementarity sky visibilities of the KM3NeT and the IceCube observatories, together with the high duty cycle of the detectors.


\acknowledgments

The authors acknowledge the financial support of the funding agencies: Czech Science Foundation (GAČR 24-12702S); Agence Nationale de la Recherche (contract ANR-15-CE31-0020), Centre National de la Recherche Scientifique (CNRS), Commission Europ\'eenne (FEDER fund and Marie Curie Program), LabEx UnivEarthS (ANR-10-LABX-0023 and ANR-18-IDEX-0001), Paris \^Ile-de-France Region, France; Shota Rustaveli National Science Foundation of Georgia (SRNSFG, FR-22-13708), Georgia; The General Secretariat of Research and Innovation (GSRI), Greece; Istituto Nazionale di Fisica Nucleare (INFN) and Ministero dell’Universit{\`a} e della Ricerca (MUR), through PRIN 2022 program (Grant PANTHEON 2022E2J4RK, Next Generation EU) and PON R\&I program (Avviso n. 424 del 28 febbraio 2018, Progetto PACK-PIR01 00021), Italy; A. De Benedittis, R. Del Burgo, W. Idrissi Ibnsalih, A. Nayerhoda, G. Papalashvili, I. C. Rea, S. Santanastaso, A. Simonelli have been supported by the Italian Ministero dell'Universit{\`a} e della Ricerca (MUR), Progetto CIR01 00021 (Avviso n. 2595 del 24 dicembre 2019); Ministry of Higher Education, Scientific Research and Innovation, Morocco, and the Arab Fund for Economic and Social Development, Kuwait; Nederlandse organisatie voor Wetenschappelijk Onderzoek (NWO), the Netherlands; The National Science Centre, Poland (2021/41/N/ST2/01177); The grant “AstroCeNT: Particle Astrophysics Science and Technology Centre”, carried out within the International Research Agendas programme of the Foundation for Polish Science financed by the European Union under the European Regional Development Fund; National Authority for Scientific Research (ANCS), Romania; Slovak Research and Development Agency under Contract No. APVV-22-0413; Ministry of Education, Research, Development and Youth of the Slovak Republic; MCIN for PID2021-124591NB-C41, -C42, -C43, funded by MCIN/AEI/10.13039/501100011033 and by “ERDF A way of making Europe”, for ASFAE/2022/014, ASFAE/2022 /023, with funding from the EU NextGenerationEU (PRTR-C17.I01), Generalitat Valenciana, and for CSIC-INFRA23013, Generalitat Valenciana for PROMETEO/2020/019, for Grant AST22\_6.2 with funding from Consejer\'{\i}a de Universidad, Investigaci\'on e Innovaci\'on and Gobierno de Espa\~na and European Union - NextGenerationEU, for CIDEGENT/2018/034, /2019/043, /2020/049, /2021/23 and for GRISOLIAP/2021/192 and EU for MSC/101025085, Spain; The European Union's Horizon 2020 Research and Innovation Programme (ChETEC-INFRA - Project no. 101008324).



\bibliographystyle{JHEP}
\bibliography{km3net_grb221009A.bib}







\end{document}

%% file: KM3NeT_authorlist_GRB221009A_JCAP.tex
\author[a]{S.~Aiello}
\author[b,ax]{A.~Albert}
\author[c]{M.~Alshamsi}
\author[d]{S. Alves Garre}
\author[f,e]{A. Ambrosone}
\author[g]{F.~Ameli}
\author[h]{M.~Andre}
\author[i]{E.~Androutsou}
\author[j]{M.~Anguita}
\author[k]{L.~Aphecetche}
% ...
\affiliation[a]{INFN, Sezione di Catania, (INFN-CT) Via Santa Sofia 64, Catania, 95123 Italy}
\affiliation[b]{Universit{\'e}~de~Strasbourg,~CNRS,~IPHC~UMR~7178,~F-67000~Strasbourg,~France}
\affiliation[c]{Aix~Marseille~Univ,~CNRS/IN2P3,~CPPM,~Marseille,~France}
\affiliation[d]{IFIC - Instituto de F{\'\i}sica Corpuscular (CSIC - Universitat de Val{\`e}ncia), c/Catedr{\'a}tico Jos{\'e} Beltr{\'a}n, 2, 46980 Paterna, Valencia, Spain}
\affiliation[e]{INFN, Sezione di Napoli, Complesso Universitario di Monte S. Angelo, Via Cintia ed. G, Napoli, 80126 Italy}
\affiliation[f]{Universit{\`a} di Napoli ``Federico II'', Dip. Scienze Fisiche ``E. Pancini'', Complesso Universitario di Monte S. Angelo, Via Cintia ed. G, Napoli, 80126 Italy}
\affiliation[g]{INFN, Sezione di Roma, Piazzale Aldo Moro 2, Roma, 00185 Italy}
\affiliation[h]{Universitat Polit{\`e}cnica de Catalunya, Laboratori d'Aplicacions Bioac{\'u}stiques, Centre Tecnol{\`o}gic de Vilanova i la Geltr{\'u}, Avda. Rambla Exposici{\'o}, s/n, Vilanova i la Geltr{\'u}, 08800 Spain}
\affiliation[i]{NCSR Demokritos, Institute of Nuclear and Particle Physics, Ag. Paraskevi Attikis, Athens, 15310 Greece}
\affiliation[j]{University of Granada, Dept.~of Computer Architecture and Technology/CITIC, 18071 Granada, Spain}
\affiliation[k]{Subatech, IMT Atlantique, IN2P3-CNRS, Nantes Universit{\'e}, 4 rue Alfred Kastler - La Chantrerie, Nantes, BP 20722 44307 France}
% ...
\affiliation[ax]{Universit{\'e} de Haute Alsace, rue des Fr{\`e}res Lumi{\`e}re, 68093 Mulhouse Cedex, France}
% ...